\documentclass[aps,prl,10pt,twocolumn,groupedaddress,notitlepage,showpacs,floatfix,longbibliography]{revtex4-1}
\pdfoutput=1
\usepackage{graphicx,graphics,epsfig,subfigure,times,bm,bbm,amssymb,amsmath,amsfonts,amsthm,mathrsfs,MnSymbol}
\usepackage{gensymb}
\usepackage[matrix,frame,arrow]{xypic}
\usepackage[pdfstartview=FitH]{hyperref}
\usepackage{subfigure}
\hypersetup{
    colorlinks=true,       
    linkcolor=red,          
   citecolor=magenta,        
    filecolor=magenta,      
    urlcolor=cyan,           
    runcolor=cyan
}
\usepackage[pdftex]{color}
\usepackage{braket}
\usepackage{enumerate}
\usepackage[normalem]{ulem}
\usepackage[usenames,dvipsnames]{xcolor}
\usepackage{multirow}
\usepackage{mathtools}
\DeclarePairedDelimiter{\ceil}{\lceil}{\rceil}

\definecolor{orange}{rgb}{1,0.5,0}

\newcommand{\bes} {\begin{subequations}}
\newcommand{\ees} {\end{subequations}}
\newcommand{\bea} {\begin{eqnarray}}
\newcommand{\eea} {\end{eqnarray}}

\definecolor{gold}{rgb}{0.85,.66,0}

\newcommand{\beq}{\begin{equation}}

\newcommand{\eeq}{\end{equation}}

\newcommand{\ignore}[1]{}

\def\s{\sigma}

\def\>{\rangle}
\def\<{\langle}

\def\s0{I}

\newcommand{\ig}[1]{}

\begin{document}
\title{Quantum networking with short-range entanglement assistance}
\author{Siddhartha Santra}
\affiliation{US Army Research Laboratory, Adelphi, Maryland 20783, USA}
\author{Vladimir S. Malinovsky}
\affiliation{US Army Research Laboratory, Adelphi, Maryland 20783, USA}
\begin{abstract}
We propose an approach to distribute high-fidelity long-range entanglement in a quantum network assisted by the entanglement supplied by auxiliary short-range paths between the network nodes. Entanglement assistance in the form of shared catalyst states is utilized to maximize the efficiency of entanglement concentration transformations over the edges of the network. The catalyst states are recycled for use in adaptive operations at the network nodes and replenished periodically using the auxiliary short-range paths. The rate of long-range entanglement distribution using such entanglement assistance is found to be significantly higher than possible without using entanglement assistance.
\end{abstract}
\maketitle


Long-range entanglement in a quantum network \cite{Wehnereaam9288} enables promising applications ranging from unconditionally secure communications \cite{qkd1120} and loophole-free tests of quantum non-locality \cite{Belltest_QR} to a network of quantum clocks \cite{qnet_clocks} and quantum-enhanced interferometry \cite{harvard_interf}. The entanglement generated initially in such networks is short-range, that is, in the form of entangled states on edges between neighboring nodes. Subsequently, long-range entanglement may be obtained between two remote nodes by identifying a path over the primary entangled edges and connecting them via entanglement swapping operations \cite{kirby-swapping} at the intermediate nodes such as in the quantum repeater \cite{munro_repeater} approach.

A quantum network should be seen as more than an interlaced collection of linear quantum repeaters. The latter utilize local quantum operations at the network nodes aided by classical communication (LOCC), such as filtration \cite{kwiat_filtration} or entanglement distillation \cite{pan_distillation}, to distribute entanglement over long distances. Yet, existing repeater protocols do not fully exploit the entanglement along auxiliary short-range paths between the nodes of the network that may be afforded, for example, by a dynamic topology \cite{topology_qn}, heterogenous links involving different degrees of freedom of the flying qudits \cite{qutrit_teleport}, or load-sharing in the quantum data plane of the network \cite{nanocon_kozlowski}. The inefficient utilization of short-range entanglement in a quantum network hinders high-fidelity long-range entanglement distribution where modest rates limit the quantum advantage of the potential applications.

\begin{figure}
\centering
\includegraphics[width=\columnwidth]{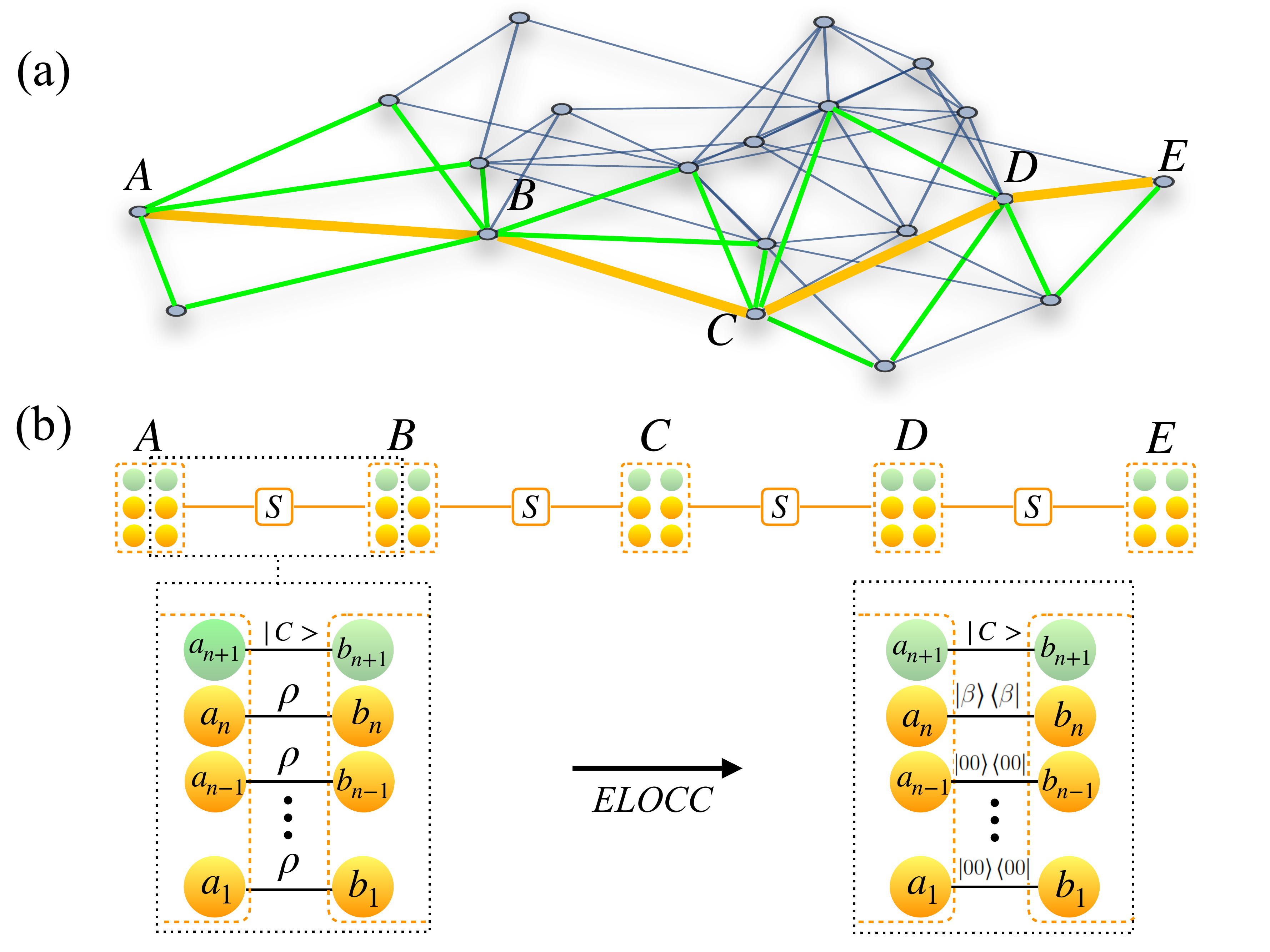}
\caption{(color online) {(a)} Schematic of a quantum network. Yellow-colored edges comprise the long-range entanglement distribution path connecting remote nodes $A$ and $E$. Green-colored edges show available auxiliary short-range paths between neighboring nodes. {(b)} An ELOCC based long-range entanglement distribution path. Neighboring half-nodes share $n$-copies of a state $\rho$ on qubit pairs $(a_i,b_i)_{i=1}^{i=n}$ (Yellow spheres) and a catalyst state $\ket{C}$ stored on the qubit pair $(a_{n+1},b_{n+1})$ (Green spheres).}
\label{fig:pic1}
\end{figure}

In this letter we propose a novel approach to efficiently utilize short-range entanglement assistance for high-fidelity long-range entanglement distribution in a quantum network. Here, auxiliary short-range paths supply catalyst states shared between network nodes that enable a  transformation of the entanglement assisted local operations and classical communication (ELOCC) class \cite{catalysis1}. This class of operations allows more general entanglement transformations in the network than LOCC. We utilize a catalytic ELOCC transformation (catalysis) to concentrate the entanglement content of primary low-fidelity states on network edges with a higher success probability compared to LOCC. Hence, a small number of the primary states are required to obtain a state with a high fidelity to a Bell state that is necessary for long-range entanglement distribution.

Catalysis is a one-shot procedure and if it succeeds the catalyst state is recovered intact along with the desired Bell state creation. The same catalyst state may be reused up to an expected number of times determined by the catalysis success probability. If the procedure fails no Bell state is obtained and the catalyst state is lost as well. For sustained long-range entanglement distribution the primary states are continuously generated over the edges along the path connecting the remote nodes in a quantum repeater-like manner. Whereas the stock of catalyst states needs to be replenished only periodically using the auxiliary short-range paths between neighboring nodes allowed by the topology of the network, see Fig.~\ref{fig:pic1}.

Our approach therefore goes beyond the linear repeater paradigm and establishes a way of quantumly aggregating the entanglement resources provided by short-range paths in a quantum network. Together with its resource-efficient and one-shot feature, this can significantly  enhance the long-range entanglement distribution rate between remote nodes in the network compared to when no auxiliary short-range entanglement assistance is available.

In catalysis, two parties at network nodes $A$ and $B$ utilize a shared entangled pure state $\ket{\mathcal{C}}$ (the catalyst) to achieve certain entanglement transformations forbidden using only LOCC \cite{Nielsen1}. Rather than being consumed the catalyst state is recovered after the transformation, i.e.,
\begin{align}
\sigma&\underset{LOCC}{\not\to}\rho\nonumber\\
\sigma\otimes \ket{\mathcal{C}}\bra{\mathcal{C}}&\underset{LOCC}{\to}  \rho\otimes \ket{\mathcal{C}}\bra{\mathcal{C}},
\end{align}
where $\sigma,\rho,\ket{\mathcal{C}}$ are bipartite states shared by $A$ and $B$. More generally, catalysis can increase the success probability of even some probabilistic entanglement transformations compatible with LOCC \cite{vidal1}. 
The set of possible catalysts for a desired transformation can be determined from the Schmidt coefficients of the initial and final states \cite{catalysis1}. For certain pairs of pure states, $\sigma=\ket{\psi}\bra{\psi}$ and $\rho=\ket{\phi}\bra{\phi}$,  a catalyst enabling a deterministic transformation, i.e. with $P_{\text{cat}}(I\to F)=1$, from an initial state $\ket{I}=\ket{\psi}\ket{\mathcal{C}}$ to a final state $\ket{F}=\ket{\phi}\ket{\mathcal{C}}$ can be found. This is possible iff the entanglement monotones \cite{vidal1}, $E_k(X):=1-\sum_{j=0}^{k-1}\lambda^X_k,~X=I,F$, that are functions of the ordered Schmidt coefficients, $\bar{\lambda}^X=(\lambda^X_1\geq\lambda^X_2\geq...\geq\lambda^X_d)$, of the initial and final states are non-increasing,
\begin{align}
E_{k}(I)\geq E_{k}(F) \forall k\in[1,d],
\label{majrel}
\end{align}
where $d$ is the larger among the dimensions of the initial and final state.

For some pairs of $\ket{\psi},\ket{\phi}$ no catalyst can make the transformation, $\ket{I}\to\ket{F}$, deterministic, i.e. $P(I\to F)\neq 1$, but can still increase the success probability as compared to just LQCC, i.e. $P_{\text{cat}}(I\to F)>P(\psi\to \phi)$ \cite{santra2020}. The success probability in this case is given by,
\begin{align}
P_{\text{cat}}(I\to F)=\underset{1\leq k \leq d}{\text{min}}R_k(I,F),
\label{LOCCprob}
\end{align}
where $R_k(I,F):=E_k(I)/E_k(F), k\in[1,d]$, is the ratio of the entanglement monotones of the initial and final states. In general, catalysts fulfilling the inequality in Eq.~(\ref{majrel}) or maximizing the R.H.S. of Eq.~(\ref{LOCCprob}) can be easily obtained using linear programming techniques \cite{walsh}. Catalysis has been shown to be more powerful than LOCC also for the case when $\sigma$ and $\rho$ are genuinely mixed states \cite{catalysis2} although no straightforward criteria such as the inequality in Eq.~(\ref{majrel}) exists in this case. 

Pure state probabilistic catalysis is already sufficient, however, to boost the long-range entanglement distribution rate in a quantum network with assistance from the entanglement along auxiliary short-range paths. We illustrate this in a case where both the primary and catalyst states are two-qubit entangled states although in general they may have different dimensions. We first motivate the form of the primary states generated on the network edges followed by a description of the catalysis process assuming availability of the catalyst. We then discuss how to obtain the said catalysts and demonstrate the enhancement that entanglement assistance obtained via auxiliary short-range paths can provide to the long-range entanglement distribution rate.

Consider a source, $S$, of polarization entangled photon pairs on an edge $A-B$ along a path connecting two remote nodes $A$ and $E$ in a quantum network as shown in Fig.~\ref{fig:pic1}. $S$, produces photon-pairs in the state, $\ket{\psi^+}=(\ket{HV}+\ket{VH})/\sqrt{2}$, which travel through the connecting optical fibers of length, $L_0$, to heralded quantum memories, e.g.  \cite{rempe_qm}, at $A$ and $B$. However, due to polarization-dependent losses in the fiber and other connecting elements the state vector of a single qubit in a superposition state undergoes the transformation, $\zeta_0\ket{H}+\zeta_1\ket{V}\to \zeta_0 (\sqrt{t_H}\ket{H}+\sqrt{1-t_H}\ket{0})+\zeta_1 \ket{V}$ \cite{danjones_filtration}. That is, a horizontally polarized photon gets effectively mixed with the vacuum mode at a beamsplitter with transmittivity $0<t_H<1$ whereas a vertically polarized photon suffers no relative loss. The state of the two quantum memories conditioned on heralding, which we assume occurs with probability $P_0$ is therefore $\rho=\ket{\alpha}\bra{\alpha}$ with,
\begin{align}
\ket{\alpha}&=\sqrt{\alpha}\ket{HV}+\sqrt{1-\alpha}\ket{VH},
\label{localstate}
\end{align}
\noindent where $\alpha=t^L_H/(t^L_H+t^R_H)$ and $t_H^{L,R}$ are the transmitivities of the left and the right channels relative to the source. In general, since $\alpha\neq (1/2)$, the primary entanglement is in the form of partially entangled states of fidelity, $F_{L_0}(\rho)=|\braket{\psi^+|\alpha}|^2=\frac{1}{2}+\sqrt{\alpha(1-\alpha)}<1$.

A direct utilization of the states $\rho$ of the form in Eq.~(\ref{localstate}) to obtain a long-range entangled state, $\rho_{AE}$, via entanglement swapping at intermediate nodes leads to a rapid decrease of the expected fidelity of $\rho_{AE}$. Assuming primary entanglement generation over each edge as described above, the expected fidelity of $\rho_{AE}$ approaches the distillability threshold exponentially fast with the total length, i.e., $\bra{\psi^+}\rho_{AE}\ket{\psi^+}-\frac{1}{2}\sim(\alpha(1-\alpha))^{N/2}$, where $N=L_{\text{tot}}/L_0$, is the number of edges along the path. Since much higher temporal resources are required for the entanglement manipulation of lower fidelity states to obtain a high-fidelity state the rate of directly obtaining the latter over long range is greatly suppressed. The fidelity of the entangled states over each edge therefore needs to be enhanced before the entanglement connections are made in order to achieve a reasonable long-range entanglement distribution rate.

An optimal catalytic transformation can efficiently provide Bell states over the network edges utilizing a few copies of the state $\rho$ and one copy of a non-maximally entangled catalyst, $\ket{\mathcal{C}}=\sqrt{c}\ket{HV}+\sqrt{1-c}\ket{VH}, 0.5<c<1$. For this, at least $(n+1)$-qubit quantum memories are required at each half-node so that $n$-copies of the state $\rho$ and one copy of the catalyst shared with the neighboring node are available for implementing the transformation,
\begin{align}
\rho^{\otimes n}\otimes \ket{\mathcal{C}}\bra{\mathcal{C}}\to \ket{\beta}\bra{\beta}\otimes \ket{00}\bra{00}^{\otimes(n-1)}\otimes \ket{\mathcal{C}}\bra{\mathcal{C}},
\label{entrf}
\end{align}
with $\ket{\beta}=(\ket{HV}+\ket{VH})/\sqrt{2}$. The transformation concentrates the entanglement in $n$-copies of the partially entangled state $\rho$ into a single maximally entangled state $\ket{\beta}$ in a two-step process using adaptive unitary operations and measurements dependent on the parameters $\alpha$ and $c$ and one-round of two-way classical communication between the nodes \cite{Nielsen1,vidal1} and is probabilistic in general \cite{santra2020} (see supp. mat. for details).

In step one, a temporary shared state $\ket{\gamma(n,\alpha,c)}$ is deterministically obtained via a sequence of operations on the initial joint state, $\ket{\alpha}^{\otimes n}\ket{\mathcal{C}}$, of qubits at $A$ and $B$. The entries of $\ket{\gamma(n,\alpha,c)}$ can be obtained algorithmically from the set of ratios, $R_l(I,F), l\in[1,2^{n+1}]$, and the entanglement monotones, $E_l(F)$, for the initial and final states given by the left and right hand sides of the transformation in Eq.~(\ref{entrf}), respectively. Essentially, $\ket{\gamma(n,\alpha,c)}$ is a state that maximizes the success probability in Eq.~(\ref{entrf}) achievable using only local measurements at either node.
In step two, a generalized measurement with two-outcomes is performed at one of the nodes on its portion of $\ket{\gamma(n,\alpha,c)}$. Corresponding to one of the measurement outcomes the state, $\ket{\beta}\otimes\ket{00}^{\otimes(n-1)}\otimes \ket{\mathcal{C}}$, is successfully obtained with probability, $P^{\text{max}}_{\text{cat}}(n,\alpha,c)$. The success (or failure) of the measurement is then relayed to the other node. In case of success, agents at $A$ and $B$ can utilize the shared state $\ket{\beta}$ for entanglement swapping to extend its range (if the adjacent edges on either side also report success) and the catalyst state for further local entanglement concentration operations. Whereas, in case of failure they restart the process of generating $n$-copies of $\ket{\alpha}$ and reobtaining the catalyst, $\ket{\mathcal{C}}$. 

The success probability, $P^{\text{max}}_{\text{cat}}(n,\alpha,c)$, of the transformation in Eq.~(\ref{entrf}) increases with $n$ for fixed $\alpha$ and $c$, that is, for a given initial state and choice of catalyst. However, generating and storing more copies of $\rho$ is temporally and spatially expensive. Further, the depth of the quantum circuit to implement the above steps at each network node scales as $\sim2^{n+1}$. Therefore, in a quantum network catalytic transformations that require collective manipulations of a smaller number of qubits are desirable. At least two-copies of $\rho$ are required for any catalyst to boost, $P^{\text{max}}_{\text{cat}}(n,\alpha,c)$,  compared to the optimal LOCC success probability, $P^{\text{max}}_{\text{LOCC}}(n,\alpha)=2(1-\alpha^n)$, of the transformation, $\rho^{\otimes n}\to\ket{\beta}\bra{\beta}\otimes \ket{00}\bra{00}^{\otimes(n-1)}$. Whereas, for a large number of copies, $n\geq n_*(\alpha)$, where $n_*(\alpha):= \ceil{1/\log_2(1/\alpha)},~\alpha\in[0.5,1)$, catalysis is not needed as in that case the same transformation can be achieved with certainty using LOCC without any catalyst. Accessing just a few copies of the primary entangled states in the range, $n\in[2,...,(n_*(\alpha)-1)]$, and utilizing the optimal two-qubit catalyst, $\ket{\mathcal{C_{\text{opt}}}(n,\alpha)}=\sqrt{c_0(n,\alpha)}\ket{HV}+\sqrt{1-c_0(n,\alpha)}\ket{VH}$, the probability, $P^{\text{max}}_{\text{cat}}(n,\alpha,c_0)=(1-\alpha^n)/(1-c_0(n,\alpha))$, can be significantly higher than $P^{\text{max}}_{\text{LOCC}}(n,\alpha)$. To quantify the increase in efficiency due to catalysis we define, $\eta_P(n,\alpha,c):=P^{\text{max}}_{\text{cat}}(n,\alpha,c)/P^{\text{max}}_{\text{LOCC}}(n,\alpha)$. This ratio diverges when the optimal two-qubit catalyst is used in Eq.~(\ref{entrf}) for poor quality primary entangled states, that is, $\eta_P(n,\alpha,c)\sim1/\sqrt{n}\sqrt{(1-\alpha)}$, for $c=c_0(n,\alpha)$ and as $\alpha\to1$.

The catalyst state itself can be obtained from the resources provided by the primary and auxiliary paths between the relevant nodes ($A,B$ in our case). It may be obtained deterministically using a LOCC procedure if the paths provide pure states that satisfy the criteria given by the inequality in Eq.~(\ref{majrel}) else it may be obtained only probabilistically. For example, in our case, it turns out that the optimal two-qubit catalyst, $\ket{\mathcal{C_{\text{opt}}}(n,\alpha)}$, in Eq.~(\ref{entrf}) is completely determined by $\alpha$ and $n$ with its Schmidt coefficient given by, $c_0(n,\alpha)=(1+3\alpha^n-((1+3\alpha^n)^2-16\alpha^{2n})^{1/2})/4\alpha^n,~n\geq2$. Therefore, it suffices to use $n_{\text{cat}}$-copies of $\ket{\alpha}$ where, $\alpha^{n_{\text{cat}}}\leq c_0(n,\alpha)$, in order to implement, $\ket{\alpha}^{\otimes n_{\text{cat}}}\to\ket{\mathcal{C_{\text{opt}}}(n,\alpha)}$, with certainty using LOCC - if the only resources available are the primary entangled states $\rho$. Note that the minimum number of copies needed to obtain the catalyst is determined by $n$ and $\alpha$ since, $n_{\text{cat}}(n,\alpha)\geq\ceil{\log(c_0(n,\alpha))/\log(\alpha)}$.

If auxiliary paths, $i=1,...,\mathcal{N}_{\text{aux-path}}$, each supplying states, $\ket{\alpha^{(i)}}$, between the nodes  are available then, $n_{\text{cat}}^{(i)}\geq\ceil{\log(c_0(n,\alpha))/\log(\alpha^{(i)})}$, copies of the states are respectively enough to obtain a single copy of $\ket{\mathcal{C_{\text{opt}}}(n,\alpha)}$. The resources provided by distinct paths can also be combined to obtain the catalyst. For example, a subset of states with different multiplicities among the $\ket{\alpha^{(i)}}$ can provide the catalyst deterministically if their product, $\prod_i \ket{\alpha^{(i)}}^{\otimes m_i},~m_i\in\mathbb{N}$, satisfies the inequality in Eq.~(\ref{majrel}) relative to $\ket{\mathcal{C_{\text{opt}}}(n,\alpha)}$. Of course, more general catalysts (other than the optimal) \cite{santra2020} may also be used for entanglement manipulation over the network edges - the choice depends on the tradeoff between the temporal resources expended to obtain the catalyst and the entanglement distribution rate.

The average temporal resource needed for every catalysis attempt over a network edge is the statistical average of the temporal resources needed for success and failure events, i.e.,
\begin{align}
\braket{T_{L_0}}=P^{\text{max}}_{\text{cat}}\braket{T_{\text{pri}}}+(1-P^{\text{max}}_{\text{cat}})\braket{T_{\text{pri+cat}}},
\label{avtimeL0}
\end{align}
where $\braket{T_{\text{pri}}}$ and $\braket{T_{\text{pri+cat}}}$ are the average times to obtain the primary entangled states and the primary entangled states along with the catalyst respectively. We assume that the primary entanglement generation takes much longer than the local operation time at the nodes which is typically the case. The terms on the right hand side of Eq.~(\ref{avtimeL0}) also depend on several parameters as we now explain. The catalysis success probability, $P^{\text{max}}_{\text{cat}}(n,\alpha,c)$, clearly, depends on the number of copies, the primary entangled state and the choice of catalyst via $n,\alpha,c$. Next, since the time-to-success of an entanglement generation attempt is geometrically distributed we obtain, $\braket{T_{\text{pri}}}=n(T_0/P_0)$, where $T_0=2L_0/c_f$, with $c_f$ the speed of light in the fiber, and $P_0$ are the average time required and probability for primary entanglement generation.

The behavior of $\braket{T_{\text{pri+cat}}}$ depends on the availability of auxiliary short-range paths between nodes $A$ and $B$. If the catalyst states are supplied exclusively by the auxiliary paths then, $\braket{T_{\text{pri+cat}}}=\text{max}\{\braket{T_{\text{pri}}},\braket{T_{\text{cat}}}\}$, is the larger among the primary entanglement generation time and the catalyst generation time. With, $\mathcal{N}_{\text{aux-path}}$ auxiliary paths, for each of which $P_i,T_i,n_{\text{cat}}^{(i)}$ are the entanglement generation probability, entanglement generation time and number of states required to obtain the catalyst, respectively, we obtain $\braket{T_{\text{cat}}}=1/\sum_{i}(n_{\text{cat}}^{(i)}P_i/\braket{T_i})\leq 1/\mathcal{N}_{\text{aux-path}}\text{min}_i(n_{\text{cat}}^{(i)}P_i/\braket{T_i})$. In the limit of a large number of auxiliary paths, $\mathcal{N}_{\text{aux-path}}>>1$, the temporal resource needed to generate the catalyst is less than the primary entanglement generation time, $\braket{T_{\text{cat}}}\leq\braket{T_{\text{pri}}}$, therefore $\braket{T_{\text{pri+cat}}}=\braket{T_{\text{pri}}}$. This is effectively the case when the auxiliary paths can supply the catalyst faster than they are consumed due to failure events which occur with an average time of $\braket{T_{\text{pri}}}/(1-P^{\text{max}}_{\text{cat}})$. In the other limit, when there are no auxiliary paths available, $\mathcal{N}_{\text{aux-path}}=0$, in which case the temporal resource needed to obtain the catalyts is budgeted from the primary entanglement generation process. Thus, in this limit the overhead of temporal resource needed is additive, $\braket{T_{\text{pri+cat}}}=\braket{T_{\text{pri}}}+\braket{T_{\text{cat}}}=(n+n_{\text{cat}}(n,\alpha))T_0/P_0$. 

The rate of long-range entanglement distribution in a quantum network along a path where the edges use optimal catalysis for entanglement concentration can be calculated as, 
\begin{align}
R_{\text{cat}}^{(N)}(n,\alpha,c)=\frac{1}{\braket{T_{L_0}}Z^{(N)}(P^{\text{max}}_{\text{cat}}(n,\alpha,c))}
\label{catrate}
\end{align}
where $Z^{(N)}(P)=\sum_{j=1}^{N}{{N}\choose{j}}\frac{(-1)^{j+1}}{1-(1-P)^j}$, \footnote{For $P$ small, $Z^{(N)}(P)\approx1/(P(2/3)^{N-1})$, which is the case for both $P^{\text{max}}_{\text{cat}}(n,\alpha,c_0)$ and $P^{\text{max}}_{\text{LOCC}}(n,\alpha)$ as $\alpha\to1$.} is the waiting-time \cite{rate-vanloock} to obtain successful events in all of the $N$-edges that comprise the path of length, $NL_0$. On the other hand, in a quantum network where the edges utilize the optimal LOCC procedure to obtain Bell states the rate is given by $R_{\text{LOCC}}^{(N)}(n,\alpha)=1/\braket{T_{\text{pri}}}Z^{(N)}(P^{\text{max}}_{\text{LOCC}}(n,\alpha))$. Note that both the rates, $R_{\text{cat}}^{(N)}(n,\alpha,c)$ and  $R_{\text{LOCC}}^{(N)}(n,\alpha)$, are independent of the final fidelity (which is $1$) but depend on the fidelity of the primary entangled states through $\alpha$. 

\begin{figure}
\centering
\includegraphics[width=\columnwidth]{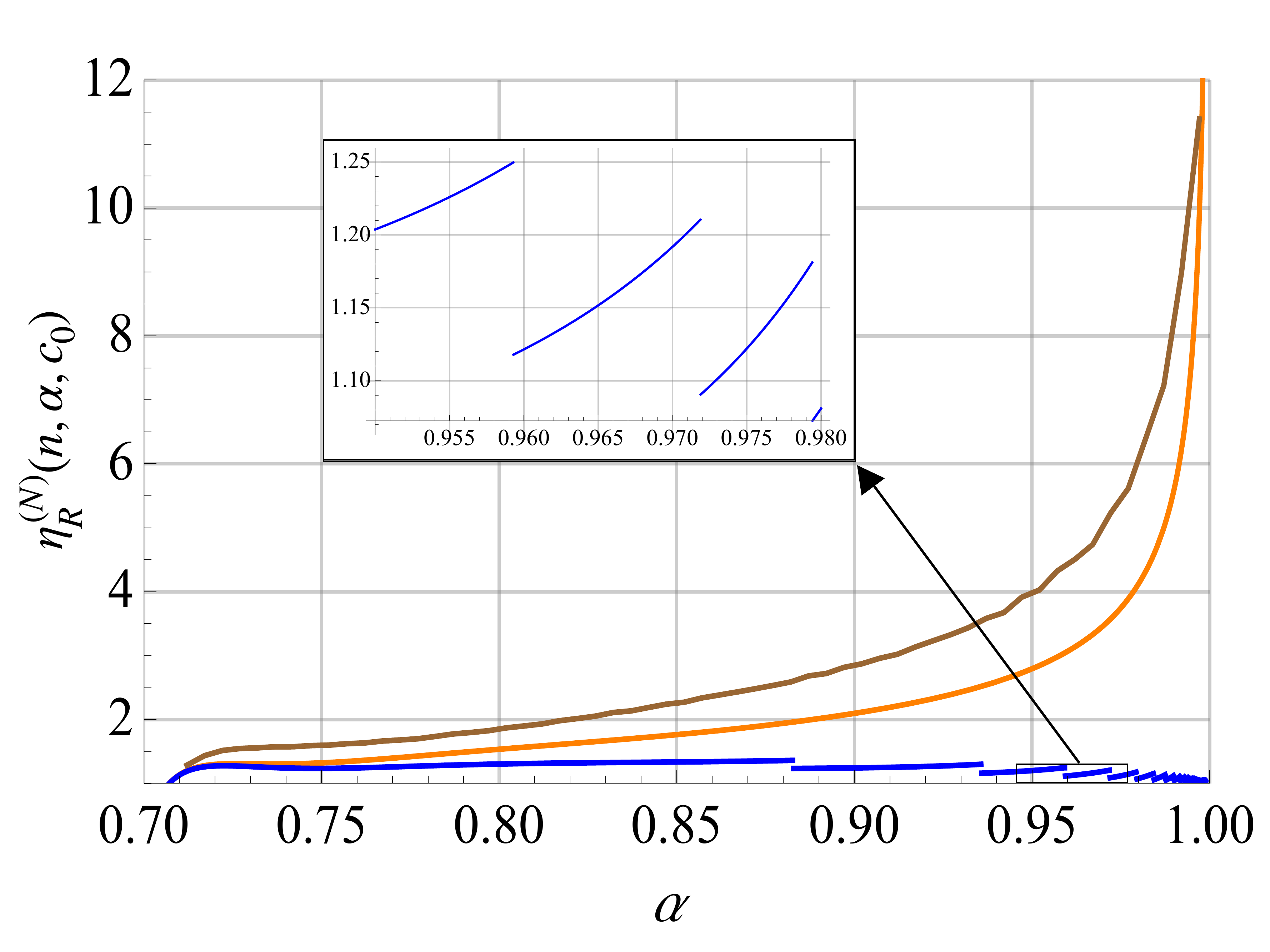}
\caption{(color online) Ratio of long-range entanglement distribution rate using optimal catalysis to the  rate using optimal LOCC for the transformation in Eq.~(\ref{entrf}) with $n=2$ on $N=2^5$ edges along a network path. Orange (Brown) curve shows the ratio using the optimal $2\times2$ ($4\times4$) catalyst when, $\mathcal{N}_{\text{aux-path}}>>1$. The discontinuous Blue curve shows the ratio with the optimal $2\times2$ catalyst for, $\mathcal{N}_{\text{aux-path}}=0$. Inset shows the nature of discontinuity - it occurs whenever the number of copies of the primary entangled state needed to obtain the catalyst jumps by 1.}
\label{fig:pic2}
\end{figure}

The ratio of the two rates, $\eta_R^{(N)}(n,\alpha,c):=R_{\text{cat}}^{(N)}(n,\alpha,c)/R_{\text{LOCC}}^{(N)}(n,\alpha)$, estimates the enhancement due to short-range entanglement assistance in the network. It is especially amplified when the primary entangled states are of poor quality and the optimal catalyst is used in the transformation in Eq.~(\ref{entrf}), that is, as $\alpha\to1$ and $c=c_0(n,\alpha)$, see Fig.~\ref{fig:pic2}. In this limit, when no auxiliary paths are available the ratio of rates approaches, $\eta_R^{(N)}(n,\alpha,c_0)\to(\braket{T_{\text{pri}}}/\braket{T_{\text{cat}}})\eta_P(n,\alpha,c_0)\approx1$. This is because the temporal overhead to generate the catalysts diminishes the advantage in rate due to the increase in the entanglement concentration success probability through catalysis. Whereas, in the same limit, $\alpha\to1$, when a large number of auxiliary paths are available the ratio diverges, $\eta_R^{(N)}(n,\alpha,c_0)\to\eta_P(n,\alpha,c_0)\sim1/\sqrt{n}\sqrt{(1-\alpha)}$ - for paths of arbitrary length. In Fig.~\ref{fig:pic2}, the marginal enhancement observed in the, $\mathcal{N}_{\text{aux-path}}=0$ and $\alpha\not\to1$ regime (Blue line), is due to the fact that catalytic entanglement concentration involves the collective manipulation of a slightly larger number of qubits in each attempt - making it a bit more efficient.

Catalysis based long-range entanglement distribution in a quantum network works particularly well under resource limited conditions - when only a small number of copies of poor quality primary entangled states are available. Its advantage is contingent upon many auxiliary short-range paths providing the catalysts. In this approach the spatial and temporal resources needed at the network nodes are constant with respect to the total length of the path and the long-range entanglement is obtained in the form of high-fidelity maximally entangled Bell states. Finally, note that the identification of suitable catalysts and design of the adaptive quantum circuit required for the catalytic transformation can be fully automated using classical computational resources at the nodes given an estimate of the primary entangled states.  

In a quantum network, catalysis can enhance or enable more general entanglement transformations beyond the pure state case since mixed state entanglement transformations can also be catalysed. Although, it does not improve the purification efficiency of Werner states \cite{werner_state}, what advantage can catalysis provide for output states of quantum channels relevant for long-range entanglement distribution, e.g., the depolarizing and dephasing channels \cite{shor_channels}? Can catalysis be useful when only partial state information is available? Finally, can catalysis be used in networks to efficiently distribute multi-party entangled states, such as GHZ or W states \cite{2d_repeaters}? 

We believe the results obtained here will stimulate research on the above mentioned questions and can facilitate developments in long-range entanglement distribution necessary for the future quantum internet.

\bibliographystyle{apsrev4-1}
\bibliography{refs-greedy1}
\end{document}